\documentclass[12pt,journal,onecolumn]{IEEEtran}

\usepackage{cite}
\usepackage{amsmath, amssymb}
\usepackage[utf8x]{inputenc}
\usepackage{alltt}

\usepackage{subfig}
\usepackage{epsfig}

\usepackage{makeidx}         
\usepackage{graphicx}        

\newtheorem{Algorithm}{Algorithm}

\newcommand{\vect} \mathbf

\begin{document}

\title{Parallel calculation of the median and order statistics on GPUs with application to robust regression}
\author{Gleb Beliakov
\\
School of Information Technology, \\Deakin University, \\221 Burwood Hwy, Burwood
3125, Australia \\ \texttt{gleb@deakin.edu.au}
}
\date{}
\maketitle


\begin{abstract}
We present and compare various approaches to a classical selection problem on Graphics Processing Units (GPUs). The selection problem consists in selecting the $k$-th smallest element from an array of size $n$, called $k$-th order statistic. We focus on calculating the median of a sample, the $n/2$-th order statistic. We introduce a new method based on  minimization of a convex  function, and show its numerical superiority when calculating the order statistics of very large arrays on GPUs. We outline an application of this approach to efficient estimation of model parameters in high breakdown robust regression.
\end{abstract}


\baselineskip=\normalbaselineskip



\section{Introduction}

We consider a classical  selection problem: calculate the $k$-order statistic of a sample $\vect x \in \Re^n$, $x_{(k)}$. Calculation of the median of a sample is a special instance of the selection problem, with $Med(\vect x)= x_{([(n+1)/2])}$, where $[t]$ denotes the integer part of $t$. Selection problem has various applications in data analysis, in particular in high-breakdown regression \cite{Maronna_book,Rousseeuw1987_book}, which is a motivating application discussed later in this paper. Selection problem is also valuable for other high-performance computing applications, such as image processing, computational aerodynamics and simulations, and it has received attention in parallel computing literature \cite{AlFuriahIEEE,Bader2004,Tiskin2009_LNCS,Saukas_conf}.

Our concern is  efficient calculation of the order statistics of very large vectors by offloading computations to GPUs. General purpose GPUs have recently become a powerful alternative to traditional CPUs, which allow one to offload various repetitive calculations to the GPU \emph{device}. GPUs have many processor cores ($p=240$ in NVIDIA's Tesla C1060 and $p=448$ in C2050), and can execute thousands of threads concurrently \cite{NVIDIA_Tesla_C1060}. The computational paradigm here is SIMT (single-instruction multiple-thread). Different streaming multiprocessors can execute different instructions, although threads within  groups (so called \emph{warp}s) are executed in SIMD (single instruction multiple data) fashion. The peak performance is achieved when there are no, or almost no branching due to conditional statements.

The selection problem is related to sorting.
A naive approach to the selection problem is to sort the elements of $\vect x$ and select the desired order statistic. On CPU, the complexity of the sort operation is $O(n \log n)$ by using one of the standard sorting algorithms. Radix sort has $O(n b)$ complexity where $b$ is the number of bits in the key \cite{Sedgewick}.

An improvement over sorting is to use a selection algorithm, such as \emph{quickselect}  \cite{NR,Sedgewick}, which is based on \emph{quicksort}, and has an expected  complexity of $O(n)$. The median of medians algorithm developed in \cite{Blum1973}, also known as \emph{BFPRT} after the last names of the authors,  has $O(n)$ worst case complexity, although empirical evidence suggests that on average \emph{quickselect} is faster.

The sort operation was parallelized for GPUs. Parallel sorting algorithms include GPU quicksort \cite{Cederman2009_ACM}, GPU sample sort \cite{Leischner2010_IEEE},  bucket-merge hybrid sort \cite{Sintorn2008_JPDC},  GPU radix sort \cite{GPU_RadixSort,Merrill:Sorting:2010} and  others \cite{S.Sengupta2007,Grand2007,N.K.Govindaraju2006}. At the  time of writing, GPU radix sort, which was implemented in Trust software \cite{Thrust}, and in the CUDA 4.0 library \cite{NVIDIA_CUDA4} released in March 2011, is the most efficient GPU sort algorithm. In empirical tests it delivered performance of up to 580 million elements/sec (for 32 bit keys on Tesla C2050) \cite{GPU_RadixSort}, which was consistent with our own evaluation.
There have been attempts to parallelize selection algorithms \cite{AlFuriahIEEE,Bader2004,Tiskin2009_LNCS,Saukas_conf} for coarse-grained multicomputers, and in the the framework of bulk-synchronous parallelism paradigm, but these  methods use divergent threads of computation. To our knowledge there is no parallel version of a selection algorithm suitable for GPUs at the moment.

In this contribution we analyze various alternatives for parallel calculation of the median of a large real vector ($n>10^5$) on a GPU device, and propose a new parallel selection method. Rather than attempting to parallelize an existing serial selection algorithm, we propose an alternative method, based on minimization of a convex cost function. Calculation of the cost function is performed by reduction in parallel. By using an effective  minimization algorithm, which requires only a few iterations, we design a  cost efficient approach to parallel selection.

The paper is structured as follows. In Section \ref{sec1} we outline the  approaches to calculating the median and order statistics in parallel and list our alternatives suitable for GPUs. In Section \ref{sec2} we present a new approach based on minimizing a convex function, or alternatively, on solving a nonlinear equation. Section \ref{sec_cp} presents Kelley's cutting plane algorithm, which we found to be the most efficient and stable among the optimization methods we evaluated.
Section \ref{sec_exper} is devoted to comparison and benchmarking of different alternatives for selection on GPUs. In Section \ref{sec_app} we outline two applications that require efficient selection algorithms, taken
 from the area of data analysis. Section \ref{sec6} concludes.

\section{Approaches to selection on GPU} \label{sec1}

Consider a  vector $\vect x \in \Re^n$, $n$ of order of $10^5-10^9$. We are interested in efficient calculation of its median, or an order statistic. As we mentioned in the Introduction, this can be achieved by using either sorting or selection algorithms.

Our problem has a specific feature.  The vector $\vect x$ is stored as an array of floats (single or double precision) on the GPU \emph{device}. The reason is that often the components of $\vect x$ are themselves calculated in parallel on GPU. This is the case in our motivating application.

In addition, we  consider a situation where $\vect x$ is so large that it does not fit the memory of a single GPU device nor CPU memory, and is distributed over several GPU devices.

Let us list our alternatives.
\begin{enumerate}
  \item Perform parallel sorting on GPU by any available GPU sorting algorithm. In our study we used the fastest GPU radix sort (we benchmarked several GPU sorting algorithms, and the results are consistent with the conclusions by Merrill and Grimshaw \cite{Merrill:Sorting:2010}).
  \item Transfer $\vect x$ to CPU and use \emph{quickselect} algorithm.
  \item Use \emph{quickselect} on GPU running as a single thread.
  \item Develop an alternative GPU selection algorithm.
\end{enumerate}

We note that when a large number of calculations of order statistics (in particular, the medians) of different vectors are required,  even small  gains in computational efficiency become important.

All the listed alternatives seem to be feasible and competitive. The  cost of alternative 2 includes the cost of copying data between GPU and CPU. The \emph{quickselect} algorithm is very efficient on modern CPUs, as compared to other sorting and selection methods. Alternative 3 eliminates the need to transfer the data, but at the expense of much slower execution of \emph{quickselect} on one of the GPU cores.

When the data are distributed over several GPUs, alternatives 2 and 3 are no longer feasible, and GPU sorting algorithms may require substantial modifications and will degrade in performance, since the data needs to be moved between different GPU devices. However, the nature of our alternative selection algorithm makes it applicable to multiple GPU situation.

\section{Approach based on search and optimization} \label{sec2}

It was known  to Laplace and Gauss that the arithmetic mean, the median, the mode, and some other functions are solutions to simple optimization problems  \cite{Jackson1921,Bullen2003_book,Gini1958_book}. A recent account and generalizations can be found in \cite{Yager1997_IJGS, Calvo2004_TFS,Beliakov2009_FSS}. In particular, the median is a minimizer of the following expression:
\begin{equation} \label{med_obj}
 Med(\vect x)= \arg \min_{y} f(y)= \arg \min_{y} \sum_{i=1}^n |x_i-y| .
\end{equation}

The objective is clearly non-smooth (it is piecewise linear), but it is convex, because the sum of convex functions is always convex. Convexity of the objective implies that there exists a unique minimum of the above expression, which can be easily found by a number of classical univariate optimization methods. Of course, numerical solution to problem (\ref{med_obj}) is  more expensive than \emph{quickselect} on CPUs, but taking into account that the data reside on the GPU, and that each evaluation of all the terms in the objective can be done in parallel in $O(\frac{n}{p})$ operations ($p$ is the number of processor cores), this method becomes a promising alternative.

For calculation of a $k-$th order statistic, the following objective is used \cite{Calvo2004_TFS,Beliakov2009_FSS}
\begin{equation} \label{med_objOS}
OS_k(\vect x)= \arg \min_{y} \sum_{i=1}^n u(x_i-y),
\end{equation}
where
$$
u(t)=\left\{ \begin{array}{cc}
               (n-k+\frac{1}{2})t & \mbox{ if } t\geq 0, \\
               -(k-\frac{1}{2})t & \mbox{ if } t<0.
             \end{array}
\right.
$$
The objective (\ref{med_objOS}) is also piecewise linear convex.

For the purposes of optimization, we considered the following alternatives:
1) Brent's method \cite{NR}, 2) the golden section method, 3) an analog of the Newton's method for  non-smooth functions \cite{Bagirov2002_oms}, and 4) Kelley's cutting plane method \cite{Kelley1960_siam}.

When the objective $f$ is differentiable, a minimization problem can be converted into a root-finding problem: solve $f'(y)=0$. For non-smooth functions we need to solve the following nonlinear equation
$$
\vect 0 \in \partial f(y),
$$
where $\partial f$ is the Clarke's subdifferential \cite{Demyanov1995_book, Clarke1983}. For a univariate function, Clarke's subdifferential at a point $t$ is simply a set of values $\partial f(t)=\{z \in \Re| \forall y \in \Re, f(y) \geq z(y-t)+f(t)\}$, such that any line with the slope $z \in \partial f(t)$  which passes through $(t,f(t))$, is a tangent line
to the graph of $f$. An element of the subdifferential is called a subgradient.

 We know that for the modulus function $$\partial |t| = \left\{ \begin{array}{cc}
  \{-1\} & \mbox{ if }t<0, \\
  \left[-1, 1\right] & \mbox{ if }t=0, \\
  \{1\} & \mbox{ if }t>0.
\end{array}
\right.$$
We can express $\partial f$ using  $g: \Re \to I \subseteq \Re$, where $I$ is a set of intervals on a real line,
$$
g(y)=count(x_i > y) + count(x_i < y)(-1) + count(x_i=y)[-1,1],
$$
where $count$ is the number of elements of $\vect x$ satisfying a given condition, and where the usual Minkowski sum for sets is used.

There are many root-finding algorithms, which can all be adapted to equation $0 \in g(y)$.
In particular, we adapted the classical bisection method, similar to how it was done in \cite{N.K.Govindaraju2004} but in a different context, as well as
 the method of parabolas combined with golden section, which is also known as Brent algorithm \cite{NR}. In fact this root finding algorithm is equivalent to  Brent's optimization method, however we looked at both alternatives, as implementation details and stopping criteria may play a role in their efficiency.


Evaluation of the objective in (\ref{med_obj}) requires summation. On GPUs it can be performed in parallel by  \emph{reduction}, using binary trees in $O(\frac{n}{p}+ \log p)$ operations. Reduction on GPU is efficiently implemented and relies on various hardware related aspects and strategies, like loop unrolling \cite{NVIDIA_Reduction}.  Furthermore, calculation of (\ref{med_obj}) on GPU involves only reading from (slow) global memory in coalescent blocks, but no writing when swapping the elements of $\vect x$, as in sorting algorithms (all temporary results are written into a much faster shared memory).
As for Eq. (\ref{med_objOS}), used for order statistics, it introduces only minimal branching in calculating each term of the sum. In the experiments we conducted, this did not produce any noticeable effect on the run time of the algorithm.

\section{Kelley's cutting plane method} \label{sec_cp}

The methods of function minimization and root finding we have used are classical, and are discussed in detail in several textbooks, e.g. \cite{NR}. Kelley's cutting plane method  \cite{Kelley1960_siam} is less known. In addition, as we show in the subsequent sections, it was the most stable and efficient algorithm in our study. Therefore it is worth to
 outline Kelley's algorithm in this section.

 The cutting plane method is relying heavily on the convexity of the objective, and requires calculation of both the objective $f$ and its subgradient $g$. This  method iteratively builds a lower piecewise linear approximation to the objective.

At each iteration, the algorithm evaluates the value of the objective $f$ and its subgradient $g$.
A piecewise linear lower estimate of $f$ is built using the formula
$$
h(y) = \max_{i=1,\ldots,k}\{ g(y_i)(y-y_i) + f(y_i)\},
$$
where $y_i$ are the points the objective was evaluated during the previous $k$ iterations. The minimizer of $h$ is chosen as the $y_{k+1}$. The algorithm starts with $y_1= x_{(1)}$ and $y_2=x_{(n)}$, which are evaluated on GPU in parallel using reduction. Figure \ref{figCP} illustrates these steps.

It is worth noting  that the minimizer of $h$ is always bracketed by two values $y_L$ and $y_R$, initially $y_1$ and $y_2$, and is
found explicitly by
$$
y_{k+1} = \frac{f(y_R)-f(y_L) + y_L g(y_L) - y_R g(y_R)}{g(y_L)-g(y_R)}.
$$
$y_L$ and $y_R$ are updated in the following way: if $g(y_{k+1})<0$ then $y_L \leftarrow y_{k+1}$ and if $g(y_{k+1})>0$ then
$y_R \leftarrow y_{k+1}$. If $g(y_{k+1})=0$, the solution is found.

Let us present Kelley's cutting plane algorithm for completeness.

\begin{Algorithm} [Cutting plane algorithm]

Input: $f$ - the objective function, $g$ - a subgradient of $f$, $maxit$ - the upper bound on the number of iterations.

Input/output: $y_L$, $y_R$ - the ends of the interval containing the minimizer.

Output: $y$ - minimizer found.

\begin{enumerate}
    \item [0.] $fL \leftarrow f(y_L), gL\leftarrow g(y_L)$, $fR\leftarrow f(y_R), gR\leftarrow g(y_R)$.
    \item [1.] for $i=1; i \leq maxit; i=i+1$ do:
   \begin{enumerate}
   \item[1.1]  $t\leftarrow (fR-fL+y_L*gL-y_R*gR)/(gL-gR)$.
   \item[1.2]  $ft\leftarrow f(t), gt\leftarrow g(t)$.
   \item[1.3]  If Stopping criteria then $y\leftarrow t$ and exit.
   \item[1.4]  If $gt<0$ \\
   then $y_L\leftarrow t, fL \leftarrow ft, gL\leftarrow gt.$ \\
   else $y_R\leftarrow t, fR\leftarrow ft, gR\leftarrow gt.$
    \end{enumerate}
    \item[2.] $y\leftarrow t$ and exit.
\end{enumerate}
\end{Algorithm}

The stopping criteria at step 1.3 comprise the following: $gt=0$ (the point with $0 \in \partial f(t)$ was found), $y_R-y_L \leq tolerance_f$, $|gt| \leq tolerance_g$.

The cutting plane algorithm in one dimension has  little overhead, and both $f$ and $g$ are easily computed in parallel and simultaneously on a GPU device. Once an approximation $\tilde y$ to the median is computed with the desired accuracy, a simple loop finds the exact value of the median in $O(\frac{n}{p} + \log p)$ operations on a GPU device by reduction \footnote{This is a simple loop which selects the largest element $x_i \leq \tilde y$. }.

The cutting plane algorithm  starts with calculation of $f$ and $g$ at the extremes of the range of the data (at step 0). The values $y_L=y_1= x_{(1)}$ and $y_R=y_2=x_{(n)}$ can be calculated in parallel by reduction. Let us also notice that the values of $f$ and $g$ at these points are computed by simple formulas: $g(y_L)=-n+2$, $g(y_R)=n-2$, $f(y_L) =\sum x_i - n y_L$  and $f(y_R)=n y_R - \sum x_i$. The values $y_L, y_R$ and $\sum x_i$ can be computed in a single parallel reduction operation, which  is advantageous compared to computations of $y_L,y_R, f(y_L)$ and $f(y_R)$ independently using four reductions. Therefore the complexity of the Algorithm 1 is at most $maxit+1$ parallel reductions.

To end this section, we notice that the cutting plane method takes only a few iterations to converge to an approximate solution $\tilde y$, under 30 iterations in our experiments with $n$ up to 32 million and $tolerance_f=10^{-12}$. Furthermore, we improved the runtime of the algorithm by combining it with parallel sorting as follows. First, we run the cutting plane method for very few iterations, say 5-7 iterations. After completing these iterations, we have an interval $[y_L, y_R]$ containing the mimimizer, which is the median of $\vect x$. We think of this interval as a pivot in selection algorithms. At the second stage, we select the values $x_i$ which fall within the (open) pivot interval $]y_L, y_R[$ and copy them to a smaller array $\vect z$. This can be done in parallel on GPU using \emph{copy\_if} operation. Then the elements of $\vect z$  are sorted in parallel using GPU radix sort. The median is then the $k-$th order statistic $z_{(k)}$ with $k=[\frac{n}{2}]-m$, and $m$ being the number of elements of $\vect x$ smaller than or equal to $y_L$, recorded during execution of the cutting plane method.

This way we obtain a hybrid selection algorithm, which benefits from the fact that sorting is performed on a much smaller array $\vect z$. The number of iterations of the cutting plane method is selected to maximize efficiency: the algorithm stops when the cost of its iteration outweighs the benefit of further reducing the pivot interval for faster conditional copy and sorting. This number is empirically selected for a given architecture. In our experiments we stopped the cutting plane algorithm after 7 iterations (for $n=2^{25}$), as at that time the pivot interval contained under $2^{19}$ elements, and its sorting was already very fast.

\section{Empirical study of computational efficiency} \label{sec_exper}

\subsection{Data sets}

We selected the following methods for numerical comparison: \emph{quickselect} on CPU, \emph{quickselect} on GPU as a single thread, GPU version of the radix sort \cite{GPU_RadixSort}, four methods based on minimization of (\ref{med_obj}) (Brent's, golden section,  nonsmooth quasi-Newton and cutting plane) and two methods based on solving $0 \in g(y)$ (bisection and Brent's root finding algorithm).

We randomly generated the following data sets of varying length $n \in \{8192=2^{13}, 32768=2^{15}, 131072=2^{17}, 524288=2^{19}, 2097152=2^{21}, 8388608=2^{23}, 33554432=2^{25}, 134 \times 10^6 \approx 2^{27}\}$:
\begin{enumerate}
  \item Uniform $x_i \sim U(0,1)$
  \item Normal $x_i \sim N(0,1)$
  \item Half-normal $x_i=|y_i|$ and $y_i \sim N(0,1)$
  \item Beta $x_i \sim \beta(2,5)$
  \item Mixture 1, 66.6\% of elements of $x_i$ chosen from $N(0,1)$ and 33.3\% from $N(100,1)$
  \item Mixture 2 50\% of elements of $x_i$ $+1$ chosen from $N(0,1)$ and the rest from $N(100,1)$
  \item Mixture 3 90\% of elements of $x_i$ chosen from half-normal $N(0,1)$ and the rest set to 10.
  \item Mixture 4 66.6\% of elements of $x_i$ chosen from half-normal $N(0,1)$ and 33.3\% from $N(100,1)$
  \item Mixture 5 50\% of elements of $x_i$ $+1$ chosen from half-normal $N(0,1)$ and the rest from $N(100,1)$
  \end{enumerate}

In addition, we performed experiments in which one or more components of $\vect x$ took very large values $\sim 10^9$.
The reasons for our selection of the distributions are the following. The median is scale invariant, so the parameters of $U$ and $N$ are taken without loss of generality. We wanted to test the algorithms for symmetric and asymmetric distributions, such as half-normal and beta, as well as for mixtures. The use of half-normal distributions is motivated by our background application, that of regression, which is described in the next section. If the data follow a linear model with normally distributed noise, the absolute residuals follow a half-normal distribution. If there are large outliers in the data, we will have a mixture of distributions for the residuals. We tested mixtures with both equal and unequal proportions of the components. These proportions correspond to the  proportion of the outliers in our background application. The model with a mixture of half-normal and normal is the closest to our application, and in fact this distribution was observed when we used regression residuals as the data.

\subsection{Algorithms and testing environment}

Our preliminary analysis allowed us to exclude several algorithms. 
Golden section was inferior to Brent's method. In fact, Brent's algorithm from \cite{NR} is organized in such a way that it uses the method of parabolas, and reverts to the golden section if the method of parabolas is unsuccessful, so it is always no worse than golden section. The quasi-Newton method was very unstable, and failed to converge in most cases.


Thus we performed detailed numerical comparison of seven algorithms: \emph{quickselect}, \emph{quickselect} on GPU,  GPU radix sort, Brent's method of optimization, cutting plane, bisection and Brent's method of solving nonlinear equations.

We measured the average time of each algorithm over 10 instances of each data set for every fixed $n$, and repeated calculations for each instance 10 times. We excluded the time spent on loading the data set from a file and on transferring it to the GPU device.

Our hardware was Tesla C2050 with $p=448$ cores and 3 GB of RAM, connected to a four-core Intel i7 CPU with 4 GB RAM clocked at 2.8 GHz, running Linux (Fedora 12). Before presenting the results, we mention some of the key performance parameters. Transfer of a 32M array of floats/doubles from GPU to CPU on our system takes over 230/455 ms, while transfer of a 500K array takes only 4/6.1 ms.  Radix sort on GPU takes 65ms for a 32M array of floats, but degrades to 226 ms for 32M array of doubles, which are similar values to those  reported in \cite{GPU_RadixSort}. One reduction on GPU when calculating objective (\ref{med_obj}) or its subgradient $g$ took 3ms for a 32M array of doubles (1.9 ms for floats).

In our implementation of the algorithms, we relied on Thrust library \cite{Thrust}, in particular, Thrust's implementation of radix sort, and the \emph{transform\_reduce} and \emph{copy\_if} functions.
Thrust library offers a number of high-level algorithms, such as reduction and transformations with arbitrary unary and binary operators, and automatically selects the
most appropriate configuration parameters (such as block size and number of blocks in a warp). Therefore we do not report experiments with these parameters, assuming that the optimal choice was made by Thrust.

We outline the program code used to calculate the the objective $f$ and its subgradient $g$ needed by Algorithm 1 in Figure \ref{f_sort1}. The actual implementation of our algorithm called \texttt{cp\_select} library, can be downloaded from \texttt{http://www.deakin.edu.au/$\sim$gleb/cp\_select.html}.
We want to stress the simplicity of the algorithm, which takes only a few lines of code relying on a high-level interface to standard algorithms, such as reduction and conditional copy, implemented in the Thrust library.

\begin{figure}[h!]
\renewcommand{\baselinestretch}{1}
 \begin{alltt}
\begin{minipage}{12cm}\small
typedef thrust::pair<double , unsigned int> Mypair;

template <typename T, typename T1>
struct abs_diff : public thrust::unary_function<T,T1>
\{
	const double y;
	abs_diff(double _a) : y(_a) \{\}
    __host__ __device__
        T1 operator()(const T& a) \{
         if(y  >= a) return Mypair(y-a,1);
                else return Mypair(a-y,0);
         \}
\};
template <typename T>
struct plusplus : public thrust::binary_function<T,T,T>
\{
    __host__ __device__
    T operator()(const T& a, const T& b)
    \{ return Mypair(a.first+b.first,a.second+b.second);  \}
\};

void Objective(int n, double t, double * f, double * df)
\{
/* calculates the values of the objective and its subgradient */
     abs_diff<double , Mypair>  unary_op(t);
     plusplus<Mypair>           binary_op;

    Mypair initpair(0.0,0);
    Mypair result =
           thrust::transform_reduce(data.begin(), data.end(),
                unary_op,  initpair, binary_op);
    *df = 2.0*result.second-n;
    *f  = result.first;
\}

void SortZ(double * result, double  L, double  R, int index)
/* Copies the data satisfying L<data[i]<R into Z and returns the n/2-index
   order statistic of Z after sorting. */
\{	
    inside_interval pred(L, R);
    DoubleIterator endZ =
          thrust::copy_if(data.begin(),data.end(),Z.begin(),pred);

    thrust::sort(Z.begin(),endZ);
    *result=Z[n/2-index];
\}

\end{minipage}
\end{alltt}
\caption{A fragment of  C code for evaluation of the objective $f(t)$ and its subgradient $g(t)$ in parallel by using Thrust \texttt{transform\_reduce} function. The function \texttt{Objective} is called by Algorithm 1, and once it terminates after $maxit$ iterations, the function \texttt{SortZ} calculates the median of $\vect x$ by sorting  array $\vect z$ containing the data in the pivot interval $[L,R]$.}\label{f_sort1}
\renewcommand{\baselinestretch}{2}
\end{figure}


\subsection{Empirical results}

\begin{figure}[htb]
  \centering
  {\includegraphics[width=0.95\textwidth]{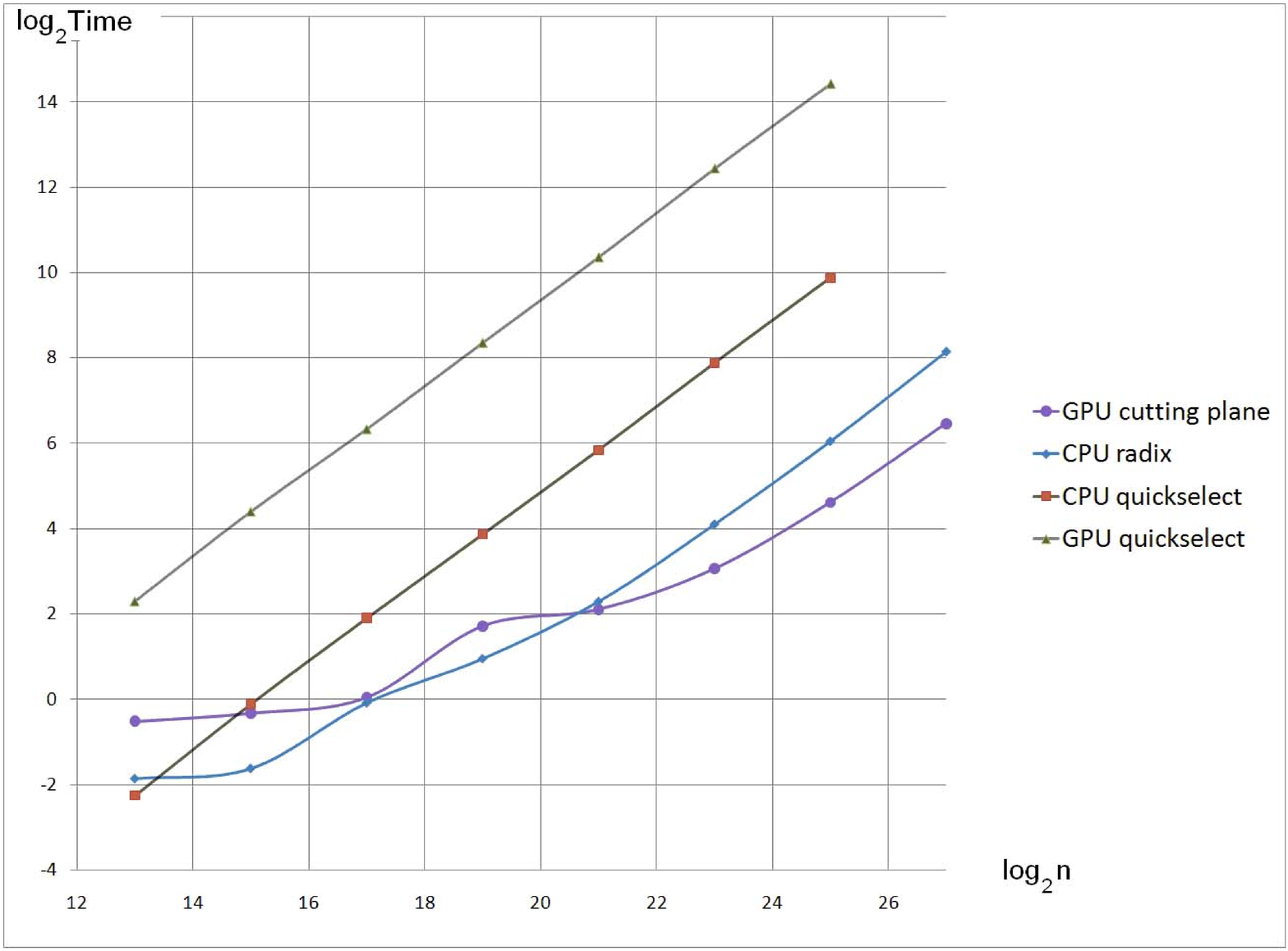}}
    \caption{Average time (in milliseconds, on a logarithmic scale) used by different methods to calculate the median, for data type \texttt{float}.}
  \label{fig2a}
\end{figure}

\begin{figure}[htb]
  \centering
  {\includegraphics[width=0.95\textwidth]{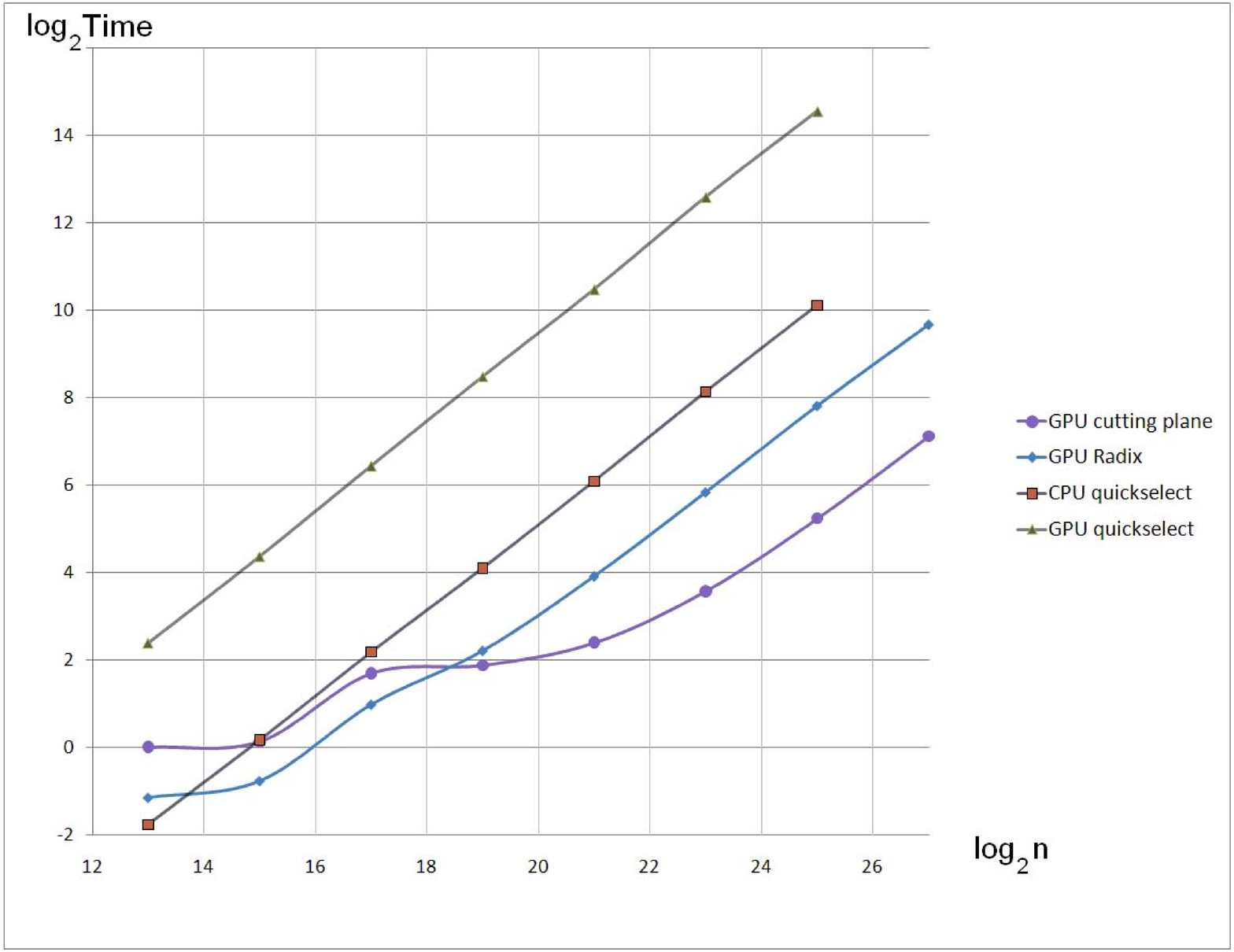}}
    \caption{Average time (in milliseconds, on a logarithmic scale) used by different methods to calculate the median, for data type \texttt{double}.}
  \label{fig2b}
\end{figure}

\begin{figure}[htb]
  \centering
  {\includegraphics[width=0.95\textwidth]{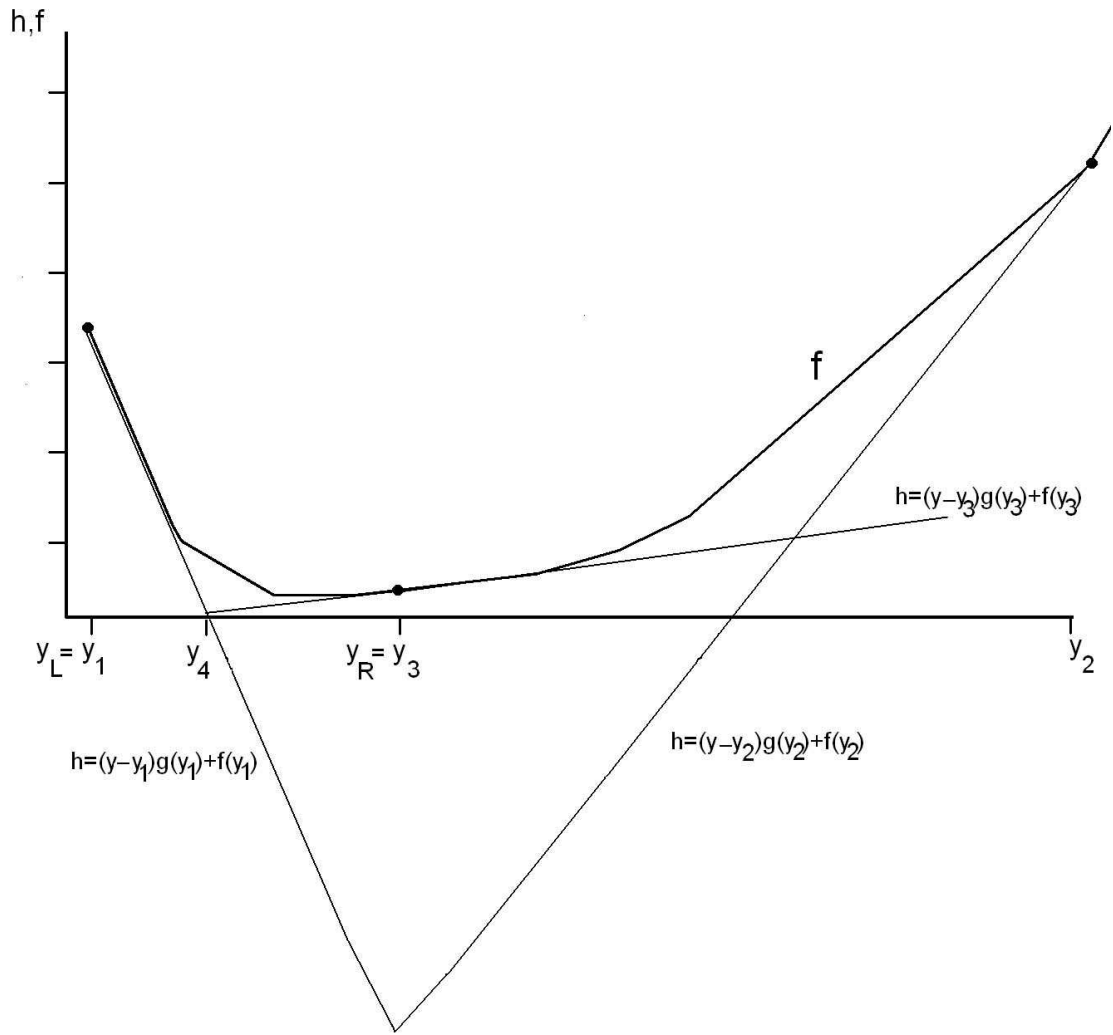}}
    \caption{Illustration of the cutting plane algorithm. The thick solid line is the graph of the objective $f$.The minimizer of $f$ is always bracketed by $y_L$ and $y_R$, which are updated at every iteration. The point of intersection of two lines tangent to the graph of $f$ yields the value $y_{k+1}$ of the new iteration.}
  \label{figCP}
\end{figure}

We present the results of our numerical experiments in Tables \ref{tab1} and \ref{tab2} for single and double precision respectively, and graphically in Figures \ref{fig2a},\ref{fig2b} (the plot is the log-log plot). We present the average results for all distributions, because we did not observe any significant deviations from the average values for  different distributions of the data. In fact, for the most efficient GPU radix sort and the proposed method, CPU times were very consistent: the largest and the smallest CPU time were within less than 5\% from the average values. For \emph{quickselect} (both CPU and GPU versions) the variations were larger, but we did not observe any pattern related to the type of the distribution, only variations between the instances of data from the same distribution.

We observe that the use of \emph{quickselect} on the CPU was more efficient only for the smallest data set tested $n=8192$, and that for larger data sets  GPU radix sort was faster by a factor of thirteen and five for single and double precision respectively. Both copying the data and actual selection algorithm on CPU have contributed to the cost of \emph{quickselect} on  CPU. We see from the tables that the time spent on just copying the data to CPU was larger than that of GPU radix sort starting from $n\geq 2^{15}$, and the same was true for the time used by \emph{quickselect} algorithm itself.
Execution of \emph{quickselect} on GPU in a single thread was worse than the remaining alternatives by a very large factor.

GPU radix sort was the most efficient method for up to $n=2^{21}$. At that point the methods based on minimization outperformed GPU radix sort and consistently remained more efficient for large data sets. It looks from the plots of the logarithms of CPU time, that starting from $n=2^{23}$ the CPU time behaved like $O(n)$ for all methods, and the difference was due to a constant factor. The difference in performance of GPU radix sort and the proposed method was more pronounced for double precision values, as one would have expected, because the performance of the radix sort depends on the number of bits in the key.

Among the methods of minimization/root finding, bisection was the slowest. Brent's root finding algorithm delivered similar performance to the cutting plane method, but its performance degraded when data contained very large outliers, as explained below.


 The difference in CPU time between the fastest existing alternative GPU radix sort and our cutting plane method is almost six-fold for large arrays $n=2^{27}$ (double) and three-fold (single precision).

\renewcommand{\baselinestretch}{1}
\begin{table}[!hp]
\begin{center}
\caption{The mean time (ms) for each method used to calculate the median: data type  \textit{float}. The best timing is in boldface. } \label{tab1}
\begin{tabular}{|l|r|r|r|r|r|r|r|r|}
\hline
& \multicolumn{8}{|c|}{Size of the data set $n$}\\
Method  &  8192 & 32768 &131072 &524288 &2097152 &8388608 &33554432 & $134 \times 10^6 $\\
 \hline
\hline

Radix Sort (on GPU)   & 0.27 &	\textbf{0.32} &	\textbf{0.94}&	\textbf{1.93}&	4.87&	 17.07&	65.68&	 282.10 \\
\hline
Quickselect (on CPU)  &\textbf{ 0.21} &	0.93&	3.75&	14.62&	57.49&	235.85&	940.1	& \\
  \quad including: &&&&&&&& \\
   \quad - copy to CPU  & 0.1 &	0.3&	1.1&	4&	14.5	&58.5	&232&
\\

  \quad- algorithm & 0.11&	0.63	&2.65	&10.62	&42.99&	177.35&	708.1&\\
  \hline
Quickselect (on GPU)  &4.88	&21.00 &	80.26&	326.59&	1311.60	&5530&	21951& \\
\hline 	\hline
Cutting Plane (total) & 0.70 &	0.80	&1.04	&3.29	&\textbf{4.32}&	\textbf{8.34}	 &\textbf{24.48}&	 \textbf{88.00} \\
  \quad including: &&&&&&&& \\
   \quad - CP iterations&  0.29	&0.32	&0.49&	2.40&	2.90&	4.80&	14.20&	62.00\\
  \quad - copy\_if& 0.19&	0.22&	0.25&	0.47&	0.61&	1.64&	5.98&	13.90\\
  \quad - Radix sort of $\vect z$ &0.22&	0.26&	0.30&	0.42&	0.81&	1.90&	4.30&	 12.10\\
 \hline
 Bisection&0.77&1.11&1.44&3.44&10.7&51.6&216.4 &\\
 \hline
Brent's minimization&0.97&1.12&1.4&2.25&5.85&20.75&86.4 &\\
 \hline
Brent's nonlinear eqn&0.68&0.79&0.87&2.34&3.92&8.79&31.25 &\\
 \hline
\end{tabular}
\end{center}
\end{table}
\renewcommand{\baselinestretch}{2}

\renewcommand{\baselinestretch}{1}
\begin{table}[h!]
\begin{center}
\caption{The mean time (ms) for each method used to calculate the median: data type  \textit{double}. The best timing is in boldface.} \label{tab2}
\begin{tabular}{|l|r|r|r|r|r|r|r|r|}
\hline
& \multicolumn{8}{|c|}{Size of the data set $n$}\\
Method  &  8192 & 32768 &131072 &524288 &2097152 &8388608 &33554432 & $134 \times 10^6 $\\
 \hline
\hline

Radix Sort (on GPU)   & 0.45&	\textbf{0.59}&	\textbf{1.98}&	4.65&	15.17	&57.52 &	 226.39&	 820.76
\\
\hline
Quickselect (on CPU)  &\textbf{0.29}&	1.12&	4.52&	17.12&	68.11&	280.83&	1107.82&\\
  \quad including: &&&&&&&& \\
   \quad - copy to CPU  & 0.1&	0.48&	1.74	&6.1	&19.1&	64&	455&
\\
  \quad- algorithm & 0.19&	0.64&	2.78&	11.02&	49.01&	216.83&	652.8&\\
  \hline
Quickselect (on GPU)  &5.25	&20.72	&87.10&	358.58&	1429.4&	6175.6&	24056.7
& \\
\hline 	\hline
Cutting Plane (total) & 1.00&	1.10&	3.24&	\textbf{3.68}&	\textbf{5.26}	 &\textbf{11.95}&	 \textbf{37.84}	&\textbf{139.84}
 \\
  \quad including: &&&&&&&& \\
   \quad - CP iterations&  0.44	&0.5&	2.10&	2.36&	3.20&	6.80&	23.00&	94.00
\\
  \quad - copy\_if& 0.28&	0.29&	0.72	&0.83&	0.96&	2.95&	9.74&	22.84\\
  \quad - Radix sort of $\vect z$ &0.28&	0.31&	0.42&	0.49&	1.10&	2.20&	5.10&	 23.00
\\
 \hline
 Bisection&1.02&1.24&4.12&5.41&12.1&61.2&298 &\\
 \hline
Brent's minimization&1.01&1.19&3.12&4.02&5.45&26.75&92.1 &\\
 \hline
Brent's nonlinear eqn&0.97&1.04&2.81&3.25&4.49&11.14&39.2 &\\
 \hline
\end{tabular}
\end{center}
\end{table}
\renewcommand{\baselinestretch}{2}

\subsection{Discussion}

We note that unlike traditional sorting and selection algorithms, our methods based on minimization and root-finding are not sensitive to how the data are organized in the array $\vect x$ (e.g., at random,  pre-sorted, or sorted in inverse order). Indeed expression (\ref{med_obj}) is invariant with respect to permutations of the components of $\vect x$.

However, the performance of all algorithms  based on minimization or solving nonlinear equations, except the cutting plane method, drastically decreased when just one or a few components of $\vect x$ took  very large values (e.g., $10^{9}$), which could happen in our application. We do not report the CPU times, because they depend on how large was the outlier. In fact we could slow down the mentioned algorithms to any degree by taking a sufficiently large outlier.
That is, all of our methods except cutting plane, are sensitive to the values of $x_i$. The reason is the following.

\begin{figure}[htb]
  \centering
  {\includegraphics[width=0.95\textwidth]{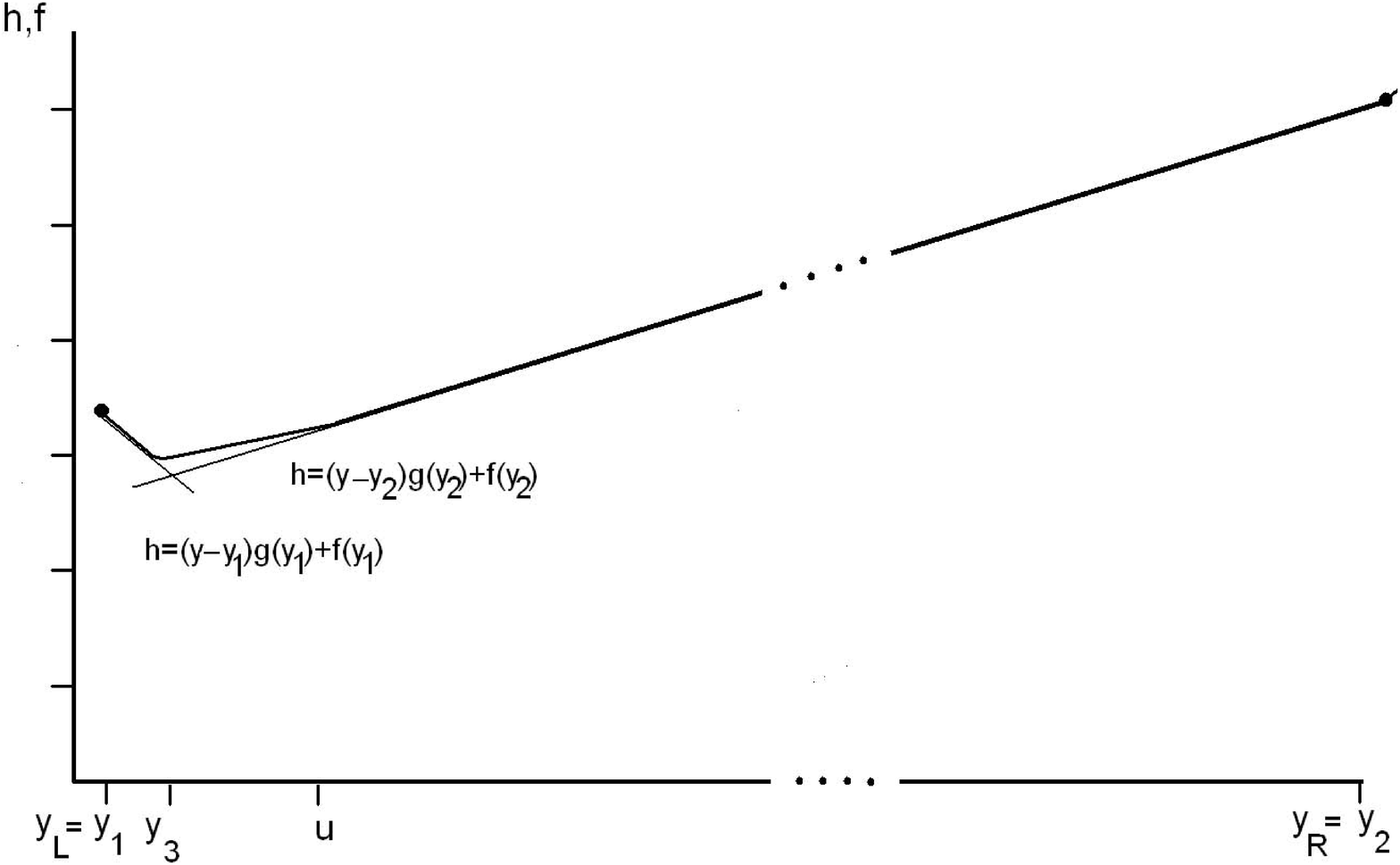}}
    \caption{Non-sensitivity of the cutting plane algorithm to extreme values of $x_i$. Even if the right end of the interval $[y_L,y_R]$ is very large, the cutting plane algorithm avoids exploring the uninteresting interval $[u,y_R]$, because on that interval $f$ coincides with $h_2(y)=(y-y_R)g(y_R)+f(y_R)$, and the next iteration  $y_3$ cannot be selected in $[u,y_R]$. In contrast, bisection makes many iterations in $[u,y_R]$, the larger $y_R$, the more iterations in $[u,y_R]$. Similarly, Brent's method attempts parabolic fit using 3 points from $[u,y_R]$, but since these points are collinear, it reverts to golden section, which again explores $[u,y_R]$.}
  \label{fig_CP1}
\end{figure}

Clearly, the number of iterations of the bisection method required to achieve a fixed accuracy is $O(\log r)$, where $r=x_{(n)} -  x_{(1)}$ is the range of the data, which can be unbounded. The same is true for the golden section optimization method, which is the counterpart of the bisection.
Both Brent's methods reverted to the slower bisection/golden section methods because the objective is linear in almost all of the range of the data when $x_i$ have such an uneven distribution. Hence the parabolic fits attempted by Brent's method are unsuccessful, see Figure \ref{fig_CP1}.

The only optimization method insensitive to very large (or small) values of $x_i$ was Kelley's cutting plane method. The reason this method was successful is that it uses not only the values of the objective but also its subgradients, as well as convexity of the  objective. The cutting plane method iteratively builds a lower piecewise linear approximation to the objective, which  also happens to be piecewise linear. In that way, just one value of the objective and its subgradient  allows the algorithm to eliminate large uninteresting  linear pieces (such as the intervals between a large outlier $x_i$ and the bulk of the data), which do not contain the minimizer.

However, even the cutting plane method had problems when some components of $\vect x$ were extremely large, of order of $10^{20}$, because of finite precision of machine arithmetic. When such a large value is accounted for in the sum (\ref{med_obj}), all the subsequent terms do not contribute to the sum because of loss of precision, even though their combined value can be sufficiently large. This is a well known effect in numerical analysis.

To circumvent this issue we used the following approach. The order statistics, and the median in particular, are invariant with respect to monotone transformations. Then we can transform the elements of $\vect x$ by applying an increasing function $F$ componentwise $F(\vect x)=(F(x_1), \ldots,F(x_n))$, calculate the median of the transformed array $med_F=Med(F(\vect x))$, and then take $med=F^{-1}(med_F)$. In our work we used transformation $F(t)=\log(1+t-x_{(1)})$. In this case the loss of accuracy in summation is eliminated.

As we mentioned in Section \ref{sec_cp}, the complexity of the cutting plane algorithm is $maxit+1$ parallel reductions, plus the time spent on conditional copy and sorting a reduced array $\vect z$. In all our experiments, the array $\vect z$ turned out to be significantly smaller than $\vect x$, typically between to 1\% and 5\% of the length of $\vect x$. Even  if, hypothetically, $\vect z$ is almost as large as $\vect x$, the complexity of the proposed method is tied to the complexity of the GPU radix sort, because the use of cutting plane  adds only a  fixed initial overhead. We were unable to design a data set which would lead to such a hypothetical case.


Finally, we will mention the situation where the data are distributed across several GPUs. Parallel sorting algorithms require transfers of data between GPUs, which has a non-negligible cost, even when such transfers do not involve the CPU, and unified virtual addressing is used, as in CUDA 4.0. In contrast, calculation of (\ref{med_obj}) and its subgradient is embarrassingly parallel, and involves reductions executed independently on different GPUs.  The partial sums from several GPUs are added together on the CPU. It works in a similar  way for multiple GPUs connected to different hosts, which are in turn connected through MPI interface, as only small portions of data need to be transferred on  few occasions (of course, we assume that the number of GPUs is not large).
 Therefore we see the approach based on minimization as the most suitable alternative to selection on multiple GPUs

\section{Application to robust regression} \label{sec_app}

Next we outline our motivating problem of high breakdown regression, the role of the median function and our GPU based algorithm. We also mention another application from data analysis which requires multiple calculations of the order statistics.

We consider the classical linear regression problem: given a set of pairs $\{(x_i,y_i)\}, i=1,\ldots,n$: $x_i \in \Re^p, y_i \in \Re$ (data), and a set of linear models $f_{\theta}: \Re^p \to \Re$ parameterized by a vector of parameters $\theta \in \Omega \subseteq \Re^{p}$ determine the parameter vector $\theta^*$, such that $f_{\theta^*}$ fits the data best. The linear models are
\begin{equation} \label{model}
y_i = x_{i1}\theta_1 + \ldots + x_{ip}\theta_{p} +\varepsilon_i , \; i=1,\ldots,n,
\end{equation}
with $x_{i p}=1$ for regression with an intercept term. $\{x_{ij}\}=X\in \Re^{n \times p}$ is the matrix of explanatory variables and $\varepsilon$ is an $n$-vector of iid random errors with zero mean and unknown variance.

 The goodness of fit is expressed in terms of either squared or absolute residuals $r_i = f_\theta(x_i)-y_i$, namely the weighted averages
$\sum_{i=1}^n w_i r_i^2$ (the least squares (LS) regression), or $\sum_{i=1}^n w_i |r_i|$ (the least absolute deviations (LAD) regression).

The LS and LAD problems are not robust to outliers: it is sufficient to take just one contaminated point with an extremely large or small value of $y_i$ or $x_{ij}$ to break the regression model. The breakdown point of the LS and LAD estimators, i.e., the proportion of data that can make the estimator's bias arbitrarily large, is 0 (see books \cite{Maronna_book,Rousseeuw1987_book}). To overcome the lack of robustness of the LS and LAD estimators, Rousseeuw \cite{Rousseeuw1984_JASA} introduced the Least Median of Squares (LMS) estimator based on the solution of
$$
\mbox{Minimize }F(\theta)=Med(r_i(\theta))^2.
$$
He also introduced the method of the least trimmed squares (LTS) \cite{Rousseeuw1984_JASA}, which is considered superior to the LMS \cite{Rousseeuw2006_DMKD,Rousseeuw1993_JASA}.
 Here the expression to be minimized is
$$
    \mbox{Minimize }F(\theta)=\sum_{i=1}^h  (r_{(i)}(\theta))^2,
$$
where the residuals are ordered in the increasing order $|r_{(1)}|\leq |r_{(2)}|\leq \ldots \leq |r_{(n)}|$, and $h=[(n+p)/2]$.

In essence, in the LMS and LTS methods, half of the sample is discarded as potential outliers, and the model is fitted to the remaining half. This makes the estimator not sensitive to contamination in up to a half of the data. Of course, the problem  is to decide which half of the data should be discarded, and this is very challenging.

Numerical computation of the LMS or LTS estimator involves multiple evaluations of $F(\theta)$. We clearly see the need in an efficient evaluation of the median in the LMS method, which is the motivation for this work. Next we also show that the median can simplify calculation of the LTS objective, even though it appears that the vector of residuals needs to be fully sorted.

We rewrite the LTS objective in this form
\begin{equation} \label{med_reg}
    \mbox{Minimize }F(\theta)=\sum_{i=1}^n  \rho( |r_{(i)}(\theta)|^2),
\end{equation}
with
$$
\rho(t) = \left\{ \begin{array}{cc}
                   1 & \mbox{ if } t < Med(\vect r) \\
                   \frac{a}{b} & \mbox{ if } t = Med(\vect r) \\
                   0 & \mbox{ otherwise}
                 \end{array}
\right.
$$
 The integers $a$ and $b$ are computed from the multiplicity of the median, i.e., the number of components of $\vect r$ equal to the median. Consider the numbers $b_L=count(|r_i|<Med(\vect r))$ and $b=count(|r_i|=Med(\vect r))$. In the LTS  method take $h=\frac{n+1}{2}$ for odd $n$, and $h=\frac{n}{2}$ for even $n$. Then $h=b_L + a$ for some $a \leq b$. Now if we take $a$ and $b$ calculated in this way in (\ref{med_reg}), the value of $F$ in (\ref{med_reg}) will be the sum of exactly $h$ smallest residuals, and will coincide with $F$ in the LTS  method.
The values of $a$ and $b$ are easily calculated once the median is known, on either CPU (in $O(n)$ time) or GPU (in $O(\frac{n}{p}+\log p)$ time) by reduction.

Hence, calculation of the median is also valuable for the LTS method, where partial sorting of the absolute residuals can be replaced with a cheaper  method based on the median.

Another application which requires multiple calculations of order statistics, and which is easily parallelized for GPUs is the k-nearest neighbor method (kNN), which is used in both regression and classification. In the case of regression, it consists in approximating a function value at $\vect x$ by $f(\vect x)=\sum_{i=1}^k w_i f_i$, where the values $f_i$ are the ordinates of the $k$ points $\vect x_1,\ldots,\vect x_k$ closest to $\vect x$ (in some metric, usually the Euclidean distance) and $w_i$ are the weights that are decreasing functions of the distances $d_i=||\vect x-\vect x_i||$. In the case of classification, the majority vote  (among the $k$ nearest neighbors) is applied to determine the predicted class of the query point $\vect x$.

The kNN method is suitable for parallelization on GPUs. The array of distances  $\vect d$, can be calculated on GPUs in parallel in $O(\frac{n}{p})$ time.
The usual approach to selecting the $k$ nearest neighbors is to sort the data according to the distance to $\vect x$. However, one can do better by using the $k$-th order statistic. Indeed, by adapting the function $\rho$ in (\ref{med_reg}), we obtain an indicator function, which returns a non-zero value for those data $(\vect x_i, f_i)$ that are no further from $\vect x$ than the $k$-order statistic $d_{(k)}$. Then the weighted sum of $k$ nearest neighbors is calculated by reduction. Thus, efficient parallel calculations of $k$-order statistics is useful in the kNN method as well.

\section{Conclusion}\label{sec6}

We considered a classical selection problem and discussed various approaches to parallel computation of the median and order statistics on GPUs.
 We compared  numerical performance of several existing and proposed methods by using a number of data sets with standard and unusual distributions, and found that the most efficient way to calculate the median is by minimizing a convex function using the cutting plane method, followed by sorting a reduced sample. This approach is between three and six times more efficient than the fastest existing method based on GPU radix sort, depending on the data type used.

 The proposed algorithm is easy to implement using high-level interface to GPU programming implemented in Thrust library, which is bundled with the newest CUDA 4.0 release. It uses a standard reduction operation, which is efficiently implemented in Thrust. Furthermore, our approach is scalable to multiple GPU devices, as only a very small number of transfers of data between GPU and CPU is required.

 We outlined two applications in data analysis which require an efficient calculation of the medians and order statistics. Both applications are easily parallelized for GPUs, and benefit from an efficient GPU-based selection algorithm.

\bibliographystyle{plain}
\bibliography{median1}
\end{document}